# Distinct Suppression Mechanisms of Superconductivity by Magnetic Domains and Spin Fluctuations in EuFe$_2$(As$_{1-x}$P$_x$)$_2$ superconductors

Mengju Yuan[1,*], Nan Zhou[2,*], Ruixia Ti[3,*], Long Zhang[1], Chenyang Zhang[4], Tian He[5], Deliu Ou[1], Jingchun Gao[1], Mingquan He[1], Aifeng Wang[1], Jun-Yi Ge [5,†], Yue Sun[2,‡], Yisheng Chai[1,§]

[1]*Low Temperature Physics Laboratory, College of Physics & Center of Quantum Materials and Devices, Chongqing University, Chongqing 401331, China*

[2]*School of Physics, Southeast University, Nanjing 211189, China*

[3]*School of Physics and Electronic Engineering, Xinxiang University, Xinxiang 453003, China*

[4]*School of Chemistry and Materials Engineering, Xinxiang University, Xinxiang 453003, China*

[5]*Materials Genome Institute, Shanghai University, Shanghai, 200444, China.*

*These authors contributed equally to this work

Contact author:

[†] Contact author: junyi_ge@t.shu.edu.cn

[‡] Contact author: sunyue@seu.edu.cn

[§] Contact author: yschai@cqu.edu.cn

Using ac composite magnetoelectric technique, we map the phase diagrams of EuFe(As$_{1-x}$P$_x$)$_2$ to resolve the interplay between superconductivity and ferromagnetism. For samples with $T_c$<$T_{FM}$, the transition to a ferromagnetic multi-domain state suppresses $H_{c2}$ through the breakdown of Jaccarino–Peter compensation and enhanced magnetic scattering from inter-domain disorder, while $H_{irr}$ is reduced due to vortex–antivortex pair nucleation at domain walls disrupting the vortex lattice. Conversely, for samples with $T_c$>$T_{FM}$, strong short-range spin correlations and phase boundaries within a multiphase coexistence regime near the triple point act as potent pair-breaking centers, leading to pronounced $H_{c2}$ suppression.

Superconductivity (SC) and ferromagnetism (FM) are traditionally viewed as antagonistic due to their opposing magnetic responses, as seen in early solid solutions like (La,Gd) and (Ce,Pr)Ru$_2$ [1,2] where the two states were mutually exclusive. Their possible coexistence was first observed in a very narrow range ($\Delta T$<1 K) of rather low temperature ($T$<1.5 K) in EuRh$_4$B$_4$ [3] and Ho$_{1.2}$Mo$_6$S$_8$ [4], later in heavy-fermion systems (e.g., UGe$_2$ [5] and UTe$_2$ with spin-triplet paring) and layered compounds RNi$_2$B$_2$C (R=Tm, Er, Ho and Dy) [6], though FM moments remained very weak or not a true ferromagnetic one. However, iron-based superconductors (IBS) [7–18], specially the recently discovered europium-containing iron pnictides [8,9,18], have transformed the field and provide a platform for robust SC–FM interplay. In particular, the EuFe$_2$(As$_{1-x}$P$_x$)$_2$ [19], a ThCr$_2$Si$_2$-type system shown in Fig. 1(a) —exhibits especially rich behavior near $x \approx 0.2$: isovalent P substitution induces strong local ferromagnetism on the Eu$^{2+}$ sublattice (~7 $\mu_B$/Eu) [20–22] while allowing SC ($T_c$~24 K) to coexist with FM ($T_{FM}$~19 K). The zero-field phase diagram from doping level $x$=0 to 1 [22]: P doping suppresses the Fe-spin-density-wave (SDW) transition, while the Eu$^{2+}$ magnetic order evolves from antiferromagnetism [low $x$, moments perpendicular to $c$-axis with possible $c$-axis canting, Fig. 1(a)] to ferromagnetism near $x\approx0.2$, where superconductivity coexists with $c$-axis-aligned FM moments (possible $ab$-plane canting). Further doping suppresses superconductivity while FM persists. Yet, this static picture does not reveal microscopic SC–FM interaction mechanisms.

To expose that interplay directly, an $H$–$T$ phase diagram near $x\approx0.2$ is essential than a purely doping-dependent one. At low fields and temperatures, coexistence of magnetic domains/domain walls and SC may yield a striped domain-Meissner state (DMS) that evolves into a domain vortex–antivortex state (DVS) upon cooling [19]. Intersections between SC transition lines (upper critical field $H_{c2}$ and irreversibility field $H_{irr}$) and magnetic boundaries (e.g., saturation field $H_s$) should then provide key coupling signatures. However, conventional resistivity and magnetization struggle to resolve both sets of transitions simultaneously, as the strong FM background masks SC signatures.

To overcome this challenge, one could employ a bulk-sensitive ac composite magnetoelectric (ME) technique that probes the ac magnetostrictive coefficient $(d\lambda/dH)_{ac}$ ($\lambda=\Delta L/L$, $L$ is the dimension of the sample). Our prior works showed that, the $(d\lambda/dH)_{ac}$ can act as a thermodynamic criterion for the vortex-lattice, capable of capturing collective vortex modes that are inaccessible to conventional static measurements. It can also exhibit pronounced dynamic magnetostrictive effect at phase transitions in ferromagnets, enabling unified extraction of SC ($H_{c2}$, $H_{irr}$) and magnetic (e.g., multi-domain ↔ spin-polarized ↔ paramagnetic) boundaries. The method has resolved $H$–$T$ diagrams in four typical type-II superconductors (Nb, $YBa_2Cu_3O_{7-x}$, $Bi_2Sr_2CaCu_2O_{8+\delta}$ and $Ba_{0.6}K_{0.4}Fe_2As_2$ [23]) and layered ferromagnets $Cr_2Ge_2Te_6$ and $Cr_2Si_2Te_6$ [24] successfully.

Here, we study two $EuFe_2(As_{1-x}P_x)_2$ crystals ($x≈0.2$) with $T_c$ below (S1) and above (S2) $T_{FM}$. Using the ac ME technique, we map their $H$–$T$ phase diagrams, precisely extract SC and magnetic transition lines, and observe clear boundary intersections. Two distinct SC suppression mechanisms are identified: In S1, the breakdown of Jaccarino–Peter compensation and inter-domain scattering suppress $H_{c2}$, while vortex–antivortex nucleation at domain walls lowers $H_{irr}$; in S2, strong pair-breaking near the triple point is driven by short-range spin correlations and phase-boundary scattering. These results demonstrate how both static magnetic domains and dynamic spin fluctuations compete with and reshape SC.

Single crystals of $EuFe_2(As_{1-x}P_x)_2$ were grown via a self-flux method [25]. The ac magnetostrictive response was measured using a composite magnetoelectric (ME) structure [left panel of Fig. 1(b)] where the crystal is bonded to a [001]-cut $0.7Pb(Mg_{1/3}Nb_{2/3})O_3$-$0.3PbTiO_3$ (PMN-PT) piezoelectric crystal using epoxy. An ac magnetic field ($H_{ac}≈0.5$–1 Oe) generated by a coil along the crystal's $c$-axis within a 14 T superconducting magnet [right panel Fig. 1(b)]. The transverse magnetostrictive response of the crystals, coupled via strain to the piezoelectric layer, induces an ac voltage $V_{ME}$ proportional to the magnetostrictive coefficient. Thus, $V_{ME}$ measured with a lock-in amplifier (OE1022 DSP) by a commercial sample stick (MultiField Tech.), yields the complex ac magnetostrictive coefficient $(d\lambda/dH)_{ac}=d\lambda'/dH+id\lambda''/dH$. Magnetic domain evolution was imaged at 2 K using a magnetic force microscopy (MFM, Attocube AttoDry2100) system. In plane resistivity ($\rho_{ab}$) and magnetization ($M$) measurements were performed using standard four-probe technique and vibration sample magnetometer in Physical Property Measurement System and Magnetic Property Measurement System (Quantum Design, QD), respectively.

We first characterized both samples using conventional probes. $\rho_{ab}$ shows metallic behavior consistent with prior reports [19] with distinct transitions [Fig. 1(c)]: S1 exhibits magnetic ordering at $T_{FM}\approx20.7$ K followed by superconducting transition at $T_c\approx14.8$ K, whereas S2 becomes superconducting first at $T_c\approx22.7$ K before ordering magnetically at $T_{FM}\approx20$ K. The zero-resistivity of S2 is achieved well below $T_c$, consistent with reported suppression from internal magnetic fields from Eu-4$f$ and Fe-3$d$ electrons [26] for a similar composition ($x$=0.27). Temperature-dependent $M(T)$ curves [Fig. 1(d)] are dominated by the Eu$^{2+}$ ferromagnetic moment. While a splitting between zero-field-cooling (ZFC) and field-cooling (FC) curves appear in both samples: in S1, the splitting is primarily magnetic, while in S2 it reflects superconducting diamagnetism, leading to a $T_{irr}\approx21.8$ K. Field-dependent $M(H)$ curves at 2 K [Fig. 1(e)] are also FM-dominated. Hysteresis persists beyond $H_s$, indicating coexisting SC, though the precise splitting field ($H_{irr}$) is hard to define. Negative initial magnetization at low fields (insets) provides further SC evidence. Together, these results directly demonstrate SC–FM coexistence in both samples, consistent with earlier reports [27]. Furthermore, MFM on S1 at 2 K shows labyrinthine domain structure at low fields that evolves into a single domain at high fields, but notably, no superconducting vortices could be resolved [see Supplemental Material FIG. S1] [28]. This confirms that conventional bulk and surface probes cannot clearly disentangle the coexisting phases.

More magnetization measurements in the FC processes show that the magnetic transition dominates the curves, obscuring the superconducting onset. As shown in Figs. 2(a) and 3(a), the magnetization increases monotonically before saturating, typical of ferromagnets. The $T_{FM}$ extracted from temperature derivative (d$M$/d$T$) peaks [Figs. 2(b) and 3(b)] become less distinct with field, shifting down below 0.4 T and upward above 0.4 T. No SC transition is discernible in $M(T)$ curves or its derivatives, highlighting how the FM background masks SC signatures.

To resolve the phase boundaries, we use the ac ME technique, where the ME voltage $V_{ME}$ is served as a proxy for $(d\lambda/dH)_{ac}$. Prior work shows that in ferromagnets: real part $d\lambda'/dH$ peaks sharply at saturation field ($H_s$) and $T_{FM}$, shows a broad hump at the spin-polarized (SP) and paramagnetic (PM) states crossover, and maximizes at the triple point (TP) where spin correlations are strongest; imaginary part $d\lambda''/dH$ can become nonzero in multidomain states [24]. In type-II superconductors, $d\lambda'/dH$ increases nearly linearly with field (plateau vs temperature) while $d\lambda''/dH \approx 0$ in the vortex-solid state; in the vortex-liquid regime, $d\lambda''/dH$ exhibits dips and $d\lambda'/dH$ decays rapidly, both vanishing near $H_{c2}$ [23]. These definitions allow unified extraction of magnetic ($H_s$, $T_{FM}$, SP–PM, TP) and superconducting ($H_{c2}$, $H_{irr}$) boundaries.

Figure 2(c) presents temperature-dependent $d\lambda'/dH$-$T$ for S1 under various fields (FC, $H//c$). Above $T_{FM}$, enhanced ferromagnetic fluctuations raise the volume magnetostriction, causing $d\lambda'/dH$ rises on cooling, peaks at $T_{FM}$, then drops sharply below it due to magnetic ordering. $T_{FM}$ values match magnetization data [Fig. 2(d)]. $d\lambda''/dH$ is near zero in the PM state but shows a pronounced dissipation peak just below $T_{FM}$, characteristic of a multidomain state [Fig. 2(d)]. $T_{FM}$ shifts down below 0.5 T and up above 0.5 T, consistent with $dM/dT$ peak positions and similar to $Cr_2Ge_2Te_6$ and $Cr_2Si_2Te_6$ [24] using the same technique. Additional weak peaks at 0.5 T and 0.6 T curves correspond to $H_s$.

Near the SC transition, $d\lambda'/dH$ drops rapidly and turns negative, while $d\lambda''/dH$ shows a dissipation dip near $T_c$, reflecting dissipative vortex-liquid dynamics. Below the irreversibility temperature $T_{irr}$, the system enters the vortex-solid phase. We define the upper and lower bounds of the $d\lambda''/dH$ dip as $T_c$ and $T_{irr}$, respectively, both shift down with increasing field. Deviations from the ideal vortex-solid response ($d\lambda'/dH$ plateau, $d\lambda''/dH \approx 0$) likely arise from vortex and ferromagnetic domains/walls interactions, consistent with reported evolution in the low-temperature DVS regime where vortex/antivortex entry and domain-period changes [29,30] alter the magnetostrictive response. Figures 2(e) and 2(f) show d$\lambda'$/d$H$ and d$\lambda''$/d$H$ in high fields (2–9 T). Over 2–30 K the signals are SC-dominated, contrasting with magnetization. At 9 T both $T_c$ and $T_{irr}$ are further suppressed, nearly extinguishing superconductivity.

Figures 3(c–f) show the (d$\lambda$/d$H$)$_{ac}$-$T$ of S2. At low fields ($\mu_0 H$<0.2 T), d$\lambda'$/d$H$ peaks at $T_{FM}$ [Fig. 3(c)] consistent with the d$M$/d$T$ peaks in Fig. 3(d). Unlike S1, the peak steep drops immediately below $T_{FM}$, reflecting the interplay of higher $T_c$ and magnetic order, the SC response is thus partially masked by spin fluctuations near $T_{FM}$. d$\lambda''$/d$H$ displays a SC-related negative dissipation dip above $T_{FM}$ that collapses rapidly below it as the system approaches the vortex-solid regime [Fig. 3(d)], indicating efficient dissipation suppression via vortex and ferromagnetic domain walls coupling. In the intermediate-field range (0.2–1 T), SC becomes dominant. $T_c$ is defined from the kink or local peak marking the onset of decrease in d$\lambda'$/d$H$, and $T_{irr}$ from the lower bound of d$\lambda''$/d$H$ dips (clearly seen in higher-field data). $T_{irr}$ decreases monotonically with field, while $T_c$ shows a non-monotonic trend (initial decrease followed by a slight increase). Below $T_{irr}$, d$\lambda'$/d$H$ still exhibits noticeable temperature dependence, likely S1 due to the vortex and ferromagnetic domains/walls interactions. Additional kinks at 0.3, 0.4 and 0.5 T curves correspond to the $H_s$. In the high-field regime (3–12 T), the response is SC-dominated across the accessible temperature range [Figs. 3(e) and 3(f)], while both $T_c$ and $T_{irr}$ shift downward with field.

Isothermal (d$\lambda$/d$H$)$_{ac}$-$H$ measurements ($H$//$c$) were also performed. For S1 up to 1 T [Figs. 4(a) and 4(b)], above 20 K (PM phase), the curves are featureless. At 20 K, a sharp d$\lambda'$/d$H$ peak marks the FM–PM transition, with strong d$\lambda''$/d$H$ dissipation indicating a multidomain state. Between 20–10 K, d$\lambda'$/d$H$ peaks at $H_s$ then decreases, reflecting Eu-spin magnetostriction, while d$\lambda''$/d$H$ shows a dissipation hump below $H_s$ from domain-wall motion and a slope change above it upon entering the SP state [Fig. 4(b)]. Below 10 K, the negative d$\lambda'$/d$H$ is dominated by the SC vortex phase, where $H_s$ appears as a local deviation from the near-linear SC response. A distinct hysteresis in $H_s$ develops between field-up and field-down sweeps, strengthening at lower temperatures and clearly seen in d$\lambda''$/d$H$ around $H_s$, but not in conventional $M(H)$ at 2 K, likely due to the broader transition features in magnetization.

Figure 4(c) shows full-span d$\lambda'$/d$H$ for S1 (0–9 T). Its amplitude rises nearly linearly with field in the vortex-solid state with growing vortex density, peaks near $H_{irr}$, then drops and vanishes close to $H_{c2}$. The $H_{irr}$–$H_{c2}$ region is marked by distinct kinks in d$\lambda''$/d$H$ [Fig. 4(d)], identifying the vortex-liquid phase. Following prior understanding [23], we define $H_{irr}$ where d$\lambda''$/d$H$ first deviates from background and $H_{c2}$ where it essentially vanishes (d$\lambda'$/d$H$ is less reliable for $H_{c2}$ due to a high-field tail). Both shift to lower fields with increasing temperature. Contrary to the expectation that dissipation is confined to the vortex-liquid regime, a pronounced, nearly field-linear response in d$\lambda''$/d$H$ below $H_{irr}$ likely arises from vortex-solid and ferromagnetic-domain interactions.

For S2 at low fields [Figs. 4(e) and 4(f)], similar to S1, the response changes from magnetic to SC dominant with $H_s$ shifting downward on warming. However, hysteresis around $H_s$ is negligible. Above $T_{FM}$, a broad positive peak shifts to higher fields (orange open symbols), indicating strong short-range spin correlations. Full-span data up to 9 T [Figs. 4(g–i)] are analyzed as in S1 for $H_{irr}$. $H_{c2}$ is defined where d$\lambda'$/d$H$ approaches zero [Fig. 4(g)], since d$\lambda''$/d$H$ has a long high-field tail [Fig. 4(i)] and is unreliable. Both shift downward with temperature, marking the vortex-liquid phase. Inside this phase, d$\lambda'$/d$H$ shows a two-stage field dependence and d$\lambda''$/d$H$ a two-dip structure—more pronounced than in S1—consistent with (d$\lambda$/d$H$)$_{ac}$-$T$ data and likely tied to multiband characteristics in Fe-based superconductors [31].

Based on (d$\lambda$/d$H$)$_{ac}$ and supported by magnetization data, we construct the $H$–$T$ phase diagrams in Fig. 5. The diagrams resolve phases of vortex-solid (V$_S$), vortex-liquid (V$_L$), SP, FM Multi-domain, and PM, with their boundary intersections providing direct signatures of SC–FM coupling. In S1 [upper panel of Fig. 5(a)], the SP→FM multi-domain transition suppresses both $H_{irr}$ and $H_{c2}$ from the high-field extrapolation just below $H_s$. $H_s$ shows pronounced hysteresis only inside the vortex-solid phase [lower panel of Fig. 5(a)], consistent with history dependence being confined to the vortex-solid regime. In S2 [Fig. 5(b)], $H_{irr}$ suppression below $H_s$ is not observed due to weak features in that region. Instead, $H_{c2}$ is significantly suppressed near the SP–PM crossover line, where strong short-range spin-spin correlations develop around the triple point (TP).

The concave curvature of $H_{irr}$ and $H_{c2}$ underscores the intricate FM–SC interplay in EuFe$_2$(As$_{1-x}$P$_x$)$_2$. In the SP phase, both samples [see Supplementary FIG. S2] display a concave $H$–$T$ dependence resembling pressure-induced superconductivity in EuFe$_2$As$_2$ [9,32,33] and Chevrel-phase compounds (Eu,$M$)Mo$_6$S$_8$ ($M$=Sn, La, etc.) [34,35]. In these systems, conduction electrons experience an internal exchange field $H_j=\beta M$ ($\beta<0$) due to antiferromagnetic coupling to localized Eu$^{2+}$ moments, described by the Jaccarino–Peter (JP) framework: $H_c(T) \approx H_c^*(0) - a[H_c(T)+H_j]^2$, with $H_c(T)$ the experimentally measured superconducting critical field, $H_c^*(0)$ the orbital critical field at $T=0$ K, and $a$ is the strength of paramagnetic pair breaking. The concave $H_c(T)$ in our $x \approx 0.2$ crystals thus implies antiferromagnetic Eu$^{2+}$ ($4f$)–Fe$^{2+}$ ($3d$) exchange [36,37]. A quantitative JP fit in the SP regime is complicated, however, because the Zeeman effect of the fully aligned Eu single domain perturbs Cooper pairs directly.

In S1, the suppression of both $H_{c2}$ and $H_{irr}$ is linked to the FM multi-domain state. Reduced net $M$ weakens JP-like compensation, allowing intrinsic paramagnetic pair-breaking to dominate $H_{c2}$ suppression, while disordered domains provide additional magnetic scattering centers. $H_{irr}$ also deviates from the high-field trend. MFM [see Supplementary FIG. S1] reveals labyrinthine domains at low fields, suggesting these walls nucleate below $T_c$ [19,20,29,30]. In Eu-based ferromagnets, antiparallel $c$-axis domains host vortex/antivortex pairs; their attractive interaction near walls [30,38,39] competes with vortex–vortex repulsion, disrupting vortex-lattice integrity and accelerating the suppression of creep activation energy at high temperatures [30], thereby suppressing $H_{irr}$. A similar $H_{irr}$ suppression is expected for S2, although multi-domain phase boundaries are not reliably tracked.

In S2, $H_{c2}$ reduction is anchored to the SP–PM crossover line, peaking near the triple point where short-range spin-spin correlations and phase coexistence are strongest. The absence of an abrupt change in $H_j$ across the crossover argues against a dominant JP-like mechanism. Moreover, the weak exchange between FeAs layers and the $Eu^{2+}$ sublattice [40,41] cannot account for the strong suppression observed. We therefore attribute the behavior to two mechanisms: (i) dynamic magnetic fluctuations from strong short-range spin correlations acting as pair-breaking centers [32,42]; and (ii) scattering at internal phase boundaries within the multiphase coexistence regime.

These results demonstrate that both static magnetic domains and dynamic spin correlations interact intricately with the vortex states, and that these interactions are probed in a highly sensitive and selective manner by $(d\lambda/dH)_{ac}$ measurements. Moreover, the pronounced hysteresis of $H_s$ in S1 within the vortex-solid phase indicates a reciprocal effect: the vortex lattice itself feeds back on, and modifies, the underlying magnetic domain state.

In summary, we have resolved the SC–FM interplay in $EuFe_2(As_{1-x}P_x)_2$ ($x\approx0.2$) using $(d\lambda/dH)_{ac}$ analysis. In samples with $T_c<T_{FM}$, the transition to a ferromagnetic multi-domain state suppresses $H_{c2}$ through the breakdown of JP-like compensation and enhanced inter-domain magnetic scattering, while $H_{irr}$ is lowered by vortex–antivortex pair nucleation at domain walls disrupting the vortex lattice. In samples with $T_c>T_{FM}$, strong pair-breaking near the triple point is driven by dynamic spin correlations and phase-boundary scattering within a multiphase coexistence regime. These results clarify the microscopic mechanisms of SC–FM coupling in $EuFe_2(As_{1-x}P_x)_2$ and establish $(d\lambda/dH)_{ac}$ as a powerful bulk probe of competing and coexisting phases in correlated electron systems.

**Acknowledgements**

We thank Jian-Wen Lu for DFT calculations on the interaction strength. We thank Gui-Wen Wang and Yan Liu at the Analytical and Testing Center of Chongqing University for technical support. This work was supported by the National Natural Science

Foundation of China (Grant Nos. 11674347, 11974065, 12174242). C. Z acknowledges the support by the National Natural Science Foundation of China (Grant No. 51802275). M. H acknowledges the support by the National Natural Science Foundation of China (Grant No. 12474141).

**DATA AVAILABILITY**

The data that support the findings of this article are not publicly available because of legal restrictions preventing unrestricted public distribution. The data are available from the authors upon reasonable request.

**Reference**

[1] B. T. Matthias, H. Suhl, and E. Corenzwit, Spin Exchange in Superconductors, Phys. Rev. Lett. **1**, 92 (1958).

[2] B. T. Matthias, H. Suhl, and E. Corenzwit, Ferromagnetic Superconductors, Phys. Rev. Lett. **1**, 449 (1958).

[3] W. A. Fertig, D. C. Johnston, L. E. DeLong, R. W. McCallum, M. B. Maple, and B. T. Matthias, Destruction of Superconductivity at the Onset of Long-Range Magnetic Order in the Compound $ErRh_4B_4$, Phys. Rev. Lett. **38**, 987 (1977).

[4] M. Ishikawa, Ø. Fischer, Destruction of superconductivity by magnetic ordering in $Ho_{1.2}Mo_6S_8$, Solid State Commun. **23**, 37 (1977).

[5] S. S. Saxena, P. Agarwal, K. Ahilan, F. M. Grosche, R. K. W. Haselwimmer, M. J. Steiner, E. Pugh, I. R. Walker, S. R. Julian, P. Monthoux *et al.*, Superconductivity on the border of itinerant-electron ferromagnetism in $UGe_2$, Nature **406**, 587 (2000).

[6] L. C. Gupta, Superconductivity and magnetism and their interplay in quaternary borocarbides $RNi_2B_2C$, Adv. Phys. **55**, 691 (2006).

[7] Y. Kamihara, T. Watanabe, M. Hirano, and H. Hosono, Iron-Based Layered Superconductor $La[O_{1-x}F_x]FeAs$ ($x = 0.05−0.12$) with $T_c = 26$ K, J. Am. Chem. Soc. **130**, 3296 (2008).

[8] Z. Ren, Q. Tao, S. Jiang, C. Feng, C. Wang, J. Dai, G. Cao, and Z. Xu,

Superconductivity Induced by Phosphorus Doping and Its Coexistence with Ferromagnetism in EuFe$_2$(As$_{0.7}$P$_{0.3}$)$_2$, Phys. Rev. Lett. **102**, 137002 (2009).

[9] T. Terashima, H. S. Suzuki, M. Tomita, M. Kimata, H. Satsukawa, A. Harada, K. Hazama, T. Matsumoto, K. Murata, and S. Uji, Pressure-induced antiferromagnetic bulk superconductor EuFe$_2$As$_2$, Phys. C **470**, S443 (2010).

[10] P. Dai, Antiferromagnetic order and spin dynamics in iron-based superconductors, Rev. Mod. Phys. **87**, 855 (2015).

[11] P. Dai, J. Hu, and E. Dagotto, Magnetism and its microscopic origin in iron-based high-temperature superconductors, Nat. Phys. **8**, 709 (2012).

[12] H. Luo, R. Zhang, M. Laver, Z. Yamani, M. Wang, X. Lu, M. Wang, Y. Chen, S. Li, S Chang *et al*, Coexistence and Competition of the Short-Range Incommensurate Antiferromagnetic Order with the Superconducting State of BaFe$_{2-x}$Ni$_x$As$_2$, Phys. Rev. Lett. **108**, 247002 (2012).

[13] Q. Huang, Y. Qiu, W. Bao, M. A. Green, J. W. Lynn, Y. C. Gasparovic, T. Wu, G. Wu, and X. H. Chen, Neutron-Diffraction Measurements of Magnetic Order and a Structural Transition in the Parent BaFe$_2$As$_2$ Compound of FeAs-Based High-Temperature Superconductors, Phys. Rev. Lett. **101**, 257003 (2008).

[14] M. Rotter, M. Tegel, and D. Johrendt, Superconductivity at 38 K in the Iron Arsenide (Ba$_{1-x}$K$_x$) Fe$_2$As$_2$, Phys. Rev. Lett. **101**, 107006 (2008).

[15] H. Luo, Z. Wang, H. Yang, P. Cheng, X. Zhu, and H.-H. Wen, Growth and characterization of A$_{1-x}$K$_x$Fe$_2$As$_2$ (A = Ba, Sr) single crystals with $x$ = 0–0.4, Supercond. Sci. Technol. **21**, 125014 (2008).

[16] Y. Chen, X. Lu, M. Wang, H. Luo, and S. Li, Systematic growth of BaFe$_{2-x}$Ni$_x$As$_2$ large crystals, Supercond. Sci. Technol. **24**, 065004 (2011).

[17] S. Jiang, Y. Luo, Z. Ren, Z. Zhu, C. Wang, X. Xu, Q. Tao, G. Cao, and Z. Xu, Metamagnetic transition in EuFe$_2$As$_2$ single crystals, New J. Phys. **11**, 025007 (2009).

[18] I. Nowik, I. Felner, Z. Ren, G. Cao, and Z. Xu, Coexistence of ferromagnetism and superconductivity: magnetization and Mössbauer studies of EuFe$_2$(As$_{1-x}$P$_x$)$_2$, J.

Phys. Condens. Matter **23**, 065701 (2011).

[19] V. S. Stolyarov, I. S. Veshchunov, S. Y. Grebenchuk, D. S. Baranov, I. A. Golovchanskiy, A. G. Shishkin, N. Zhou, Z. Shi, X. Xu, S. Pyon *et al.*, Domain Meissner state and spontaneous vortex-antivortex generation in the ferromagnetic superconductor EuFe$_2$(As$_{0.79}$P$_{0.21}$)$_2$, Sci. Adv. **4**, eaat1061 (2018).

[20] W. Jin, S. Mühlbauer, P. Bender, Y. Liu, S. Demirdis, Z. Fu, Y. Xiao, S. Nandi, G. Cao, Y. Su *et al.*, Bulk domain Meissner state in the ferromagnetic superconductor EuFe$_2$(As$_{0.8}$P$_{0.2}$)$_2$: Consequence of compromise between ferromagnetism and superconductivity, Phys. Rev. B **105**, L180504 (2022).

[21] S. Ghimire, M. Kończykowski, K. Cho, M. A. Tanatar, D. Torsello, I. S. Veshchunov, T. Tamegai, G. Ghigo, and R. Prozorov, Effect of Controlled Artificial Disorder on the Magnetic Properties of EuFe$_2$(As$_{1-x}$P$_x$)$_2$ Ferromagnetic Superconductor, Materials **14**, 3267 (2021).

[22] G. Cao, S. Xu, Z. Ren, S. Jiang, C. Feng, and Z. Xu, Superconductivity and ferromagnetism in EuFe$_2$(As$_{1-x}$P$_x$)$_2$, J. Phys. Condens. Matter **23**, 464204 (2011).

[23] P. Lu, M. Yuan, J. Zhang, Q. Gao, S. Liu, Y. Zhang, S. Shen, L. Zhang, J. Lu, X. Zhou *et al.*, Identifying vortex lattice in type-II superconductors via the dynamic magnetostrictive effect, arXiv: 2506.08873 (2025).

[24] Y. Zhang, Z. Li, J. Zhang, N. Cao, L. Zhang, Y. Li, S. Liu, X. Zhou, Y. Sun, W. Wang *et al.*, Observation of enhanced ferromagnetic spin-spin correlations at a triple point in quasi-two-dimensional magnets, Phys. Rev. B **107**, 134417 (2023).

[25] X. Xu, W. Jiao, N. Zhou, Y. Li, B. Chen, C. Cao, J. Dai, A. F. Bangura, and G. Cao, Electronic Nematicity Revealed by Torque Magnetometry in Iron Arsenide EuFe$_2$(As$_{1-x}$P$_x$)$_2$, Phys. Rev. B **89**, 104517 (2014).

[26] A. Ahmed, M. Itou, S. Xu, Z. Xu, G. Cao, Y. Sakurai, J. Penner-Hahn, and A. Deb, Competing Ferromagnetism and Superconductivity on FeAs Layers in EuFe$_2$(As$_{0.73}$P$_{0.27}$)$_2$, Phys. Rev. Lett. **105**, 207003 (2010).

[27] J. Munevar, H. Micklitz, M. Alzamora, C. Argüello, T. Goko, F. L. Ning, A. A. Aczel, T. Munsie, T. J. Williams, G. F. Chen *et al.*, Magnetism in superconducting

EuFe$_2$As$_{1.4}$P$_{0.6}$ single crystals studied by local probes, Solid State Commun. **187**, 18 (2014).

[28] See Supplemental Material for additional figures of EuFe$_2$(As$_{1-x}$P$_x$)$_2$ single crystals.

[29] S. Yu. Grebenchuk, Zh. A. Devizorova, I. A. Golovchanskiy, I. V. Shchetinin, G. Cao, A. I. Buzdin, D. Roditchev, and V. S. Stolyarov, Crossover from ferromagnetic superconductor to superconducting ferromagnet in P-doped EuFe$_2$(As$_{1–x}$P$_x$)2, Phys. Rev. B **102**, 144501 (2020).

[30] J. A. Wilcox, L. Schneider, E. Marchiori, V. Plastovets, A. Buzdin, P. Sahafi, A. Jordan, R. Budakian, T. Ren, I. Veshchunov *et al.*, Magnetically controlled vortex dynamics in a ferromagnetic superconductor, Commun. Mater. **6**, 108 (2025).

[31] E. Smørgrav, J. Smiseth, E. Babaev, and A. Sudbø, Vortex Sublattice Melting in a Two-Component Superconductor, Phys. Rev. Lett. **94**, 096401 (2005).

[32] N. Kurita, M. Kimata, K. Kodama, A. Harada, M. Tomita, H. S. Suzuki, T. Matsumoto, K. Murata, S. Uji, and T. Terashima, Upper critical field of the pressure-induced superconductor EuFe$_2$As$_2$, Phys. Rev. B **83**, 100501 (2011).

[33] T. Terashima, M. Kimata, H. Satsukawa, A. Harada, K. Hazama, S. Uji, H. S. Suzuki, T. Matsumoto, and K. Murata, EuFe$_2$As$_2$ under High Pressure: An Antiferromagnetic Bulk Superconductor, J. Phys. Soc. Jpn. **78**, 083701 (2009).

[34] O. Fischer, M. Decroux, S. Roth, R. Chevrel, and M. Sergent, Compensation of the paramagnetic effect on $H_{c2}$ by magnetic moments: 700 kG superconductors, J. Phys. C Solid State Phys. **8**, L474 (1975).

[35] M. S. Torlkachvili and M. B. Maple, Enhancement of the upper critical magnetic field in La$_{1.2-x}$Eu$_x$Mo$_6$S$_8$ compounds, Solid State Commun. **40**, 1 (1981).

[36] J. Maiwald and P. Gegenwart, Interplay of 4$f$ and 3$d$ moments in EuFe$_2$As$_2$ iron pnictides, Phys. Status Solidi B **254**, 1600150 (2017).

[37] M. Decroux, S. E. Lambert, M. S. Torikachvili, M. B. Maple, R. P. Guertin, L. D. Woolf, and R. Baillif, Observation of Bulk Superconductivity in EuMo$_6$S$_8$ under Pressure, Phys. Rev. Lett. **52**, 1563 (1984).

[38] A. A. Bespalov, A. S. Mel'nikov, and A. I. Buzdin, Clustering of vortex matter in

superconductor-ferromagnet superlattices, EPL Europhys. Lett. **110**, 37003 (2015).

[39] G. Prando, D. Torsello, S. Sanna, M. J. Graf, S. Pyon, T. Tamegai, P. Carretta, and G. Ghigo, Complex vortex-antivortex dynamics in the magnetic superconductor EuFe$_2$(As$_{0.7}$P$_{0.3}$)$_{0.2}$, Phys. Rev. B **105**, 224504 (2022).

[40] S. Nandi, W. T. Jin, Y. Xiao, Y. Su, S. Price, D. K. Shukla, J. Strempfer, H. S. Jeevan, P. Gegenwart, and Th. Brückel, Coexistence of superconductivity and ferromagnetism in P-doped EuFe$_2$As$_2$, Phys. Rev. B **89**, 014512 (2014).

[41] H. S. Jeevan, D. Kasinathan, H. Rosner, and P. Gegenwart, Interplay of antiferromagnetism, ferromagnetism, and superconductivity in EuFe$_2$(As$_{1-x}$P$_x$)$_2$ single crystals, Phys. Rev. B **83**, 054511 (2011).

[42] P. Wiecki, V. Ogloblichev, A. Pandey, D. C. Johnston, and Y. Furukawa, Coexistence of antiferromagnetic and ferromagnetic spin correlations in SrCo$_2$As$_2$ revealed by $^{59}$Co and $^{75}$As NMR, Phys. Rev. B **91**, 220406 (2015).

**Figures**

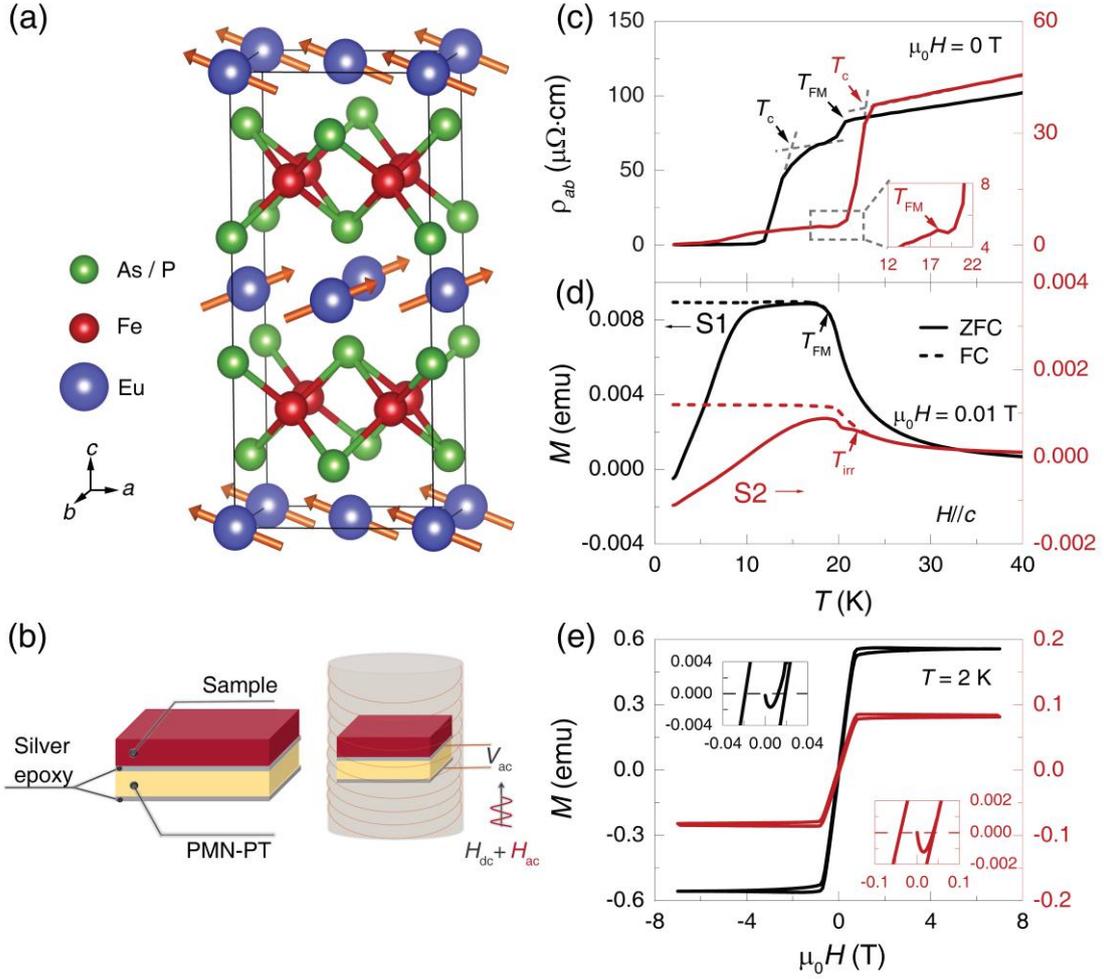

FIG. 1. (a) Crystal and magnetic structures of $EuFe_2(As_{1-x}P_x)_2$. (b) Schematic of the composite ME measurement setup. (c) In-plane resistivity vs temperature at 0 T; inset shows the low-temperature FM transition in S2. (d) Magnetization vs temperature at 0.01 T ($H//c$). (e) Magnetization vs field ($H//c$) at 2 K; inset highlights low-field SC diamagnetism.

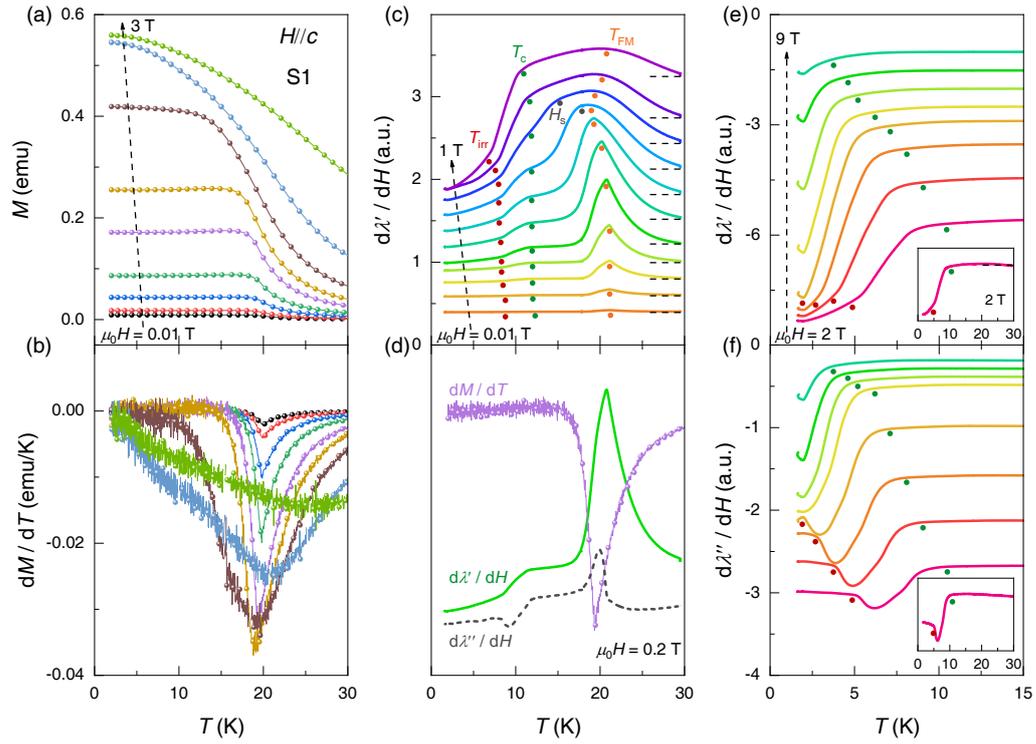

FIG. 2. S1 data: Temperature dependence of (a) $M$, (b) $dM/dT$ and (c) $d\lambda'/dH$ at low fields. (d) $dM/dT$, $d\lambda'/dH$, and $d\lambda''/dH$ at 0.2 T. (e) $d\lambda'/dH$ and (f) $d\lambda''/dH$ at high fields. Insets in (e, f) show expanded data from 2–30 K at 2 T. Traces in (c), (e), and (f) are vertically offset. Dashed lines indicate zero level of $d\lambda'/dH$.

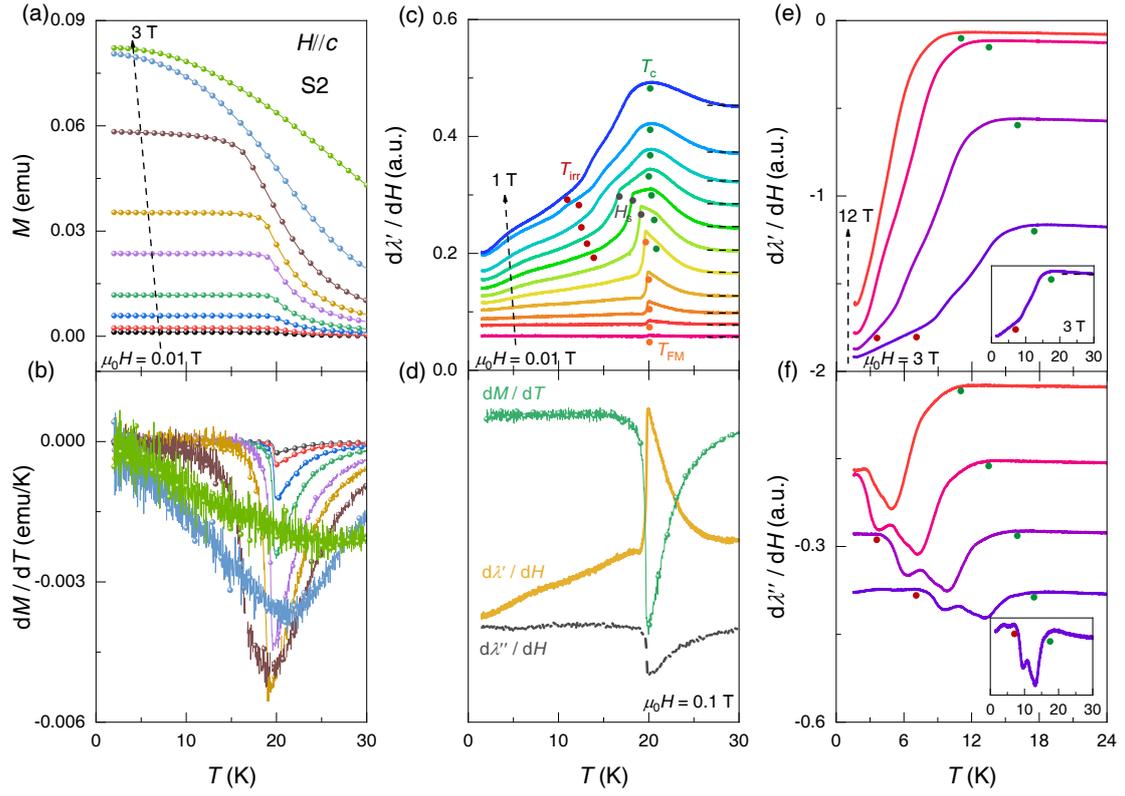

FIG. 3. S2 data: Temperature dependent (a) $M$, (b) $dM/dT$ and (c) $d\lambda'/dH$ at low fields. (d) $dM/dT$, $d\lambda'/dH$, and $d\lambda''/dH$ at 0.1 T, (e) $d\lambda'/dH$ and (f) $d\lambda''/dH$ at high fields. Insets in (e, f) show expanded data from 2–30 K at 3 T. Traces in (c), (e), and (f) are vertically offset. Dashed lines indicate zero level of $d\lambda'/dH$.

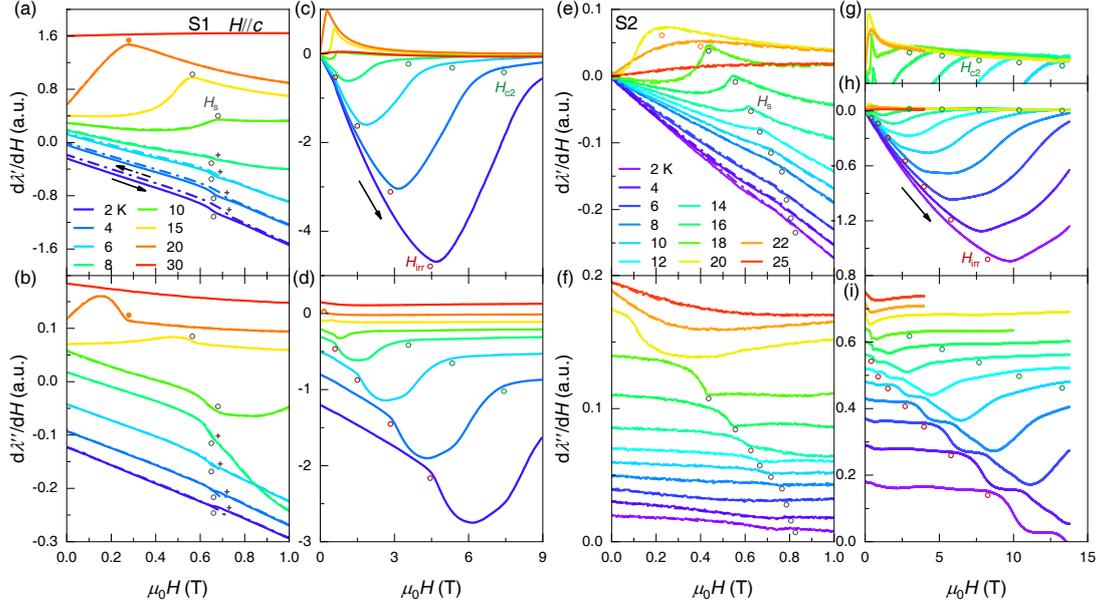

FIG. 4. Field dependent d$\lambda'$/d$H$ and d$\lambda''$/d$H$ for S1 (a–d) and S2 (e–i). (g) Magnified view of the $H_{c2}$ region from (h). Traces in (a), (b), (d), (f) and (i) are vertically offset.

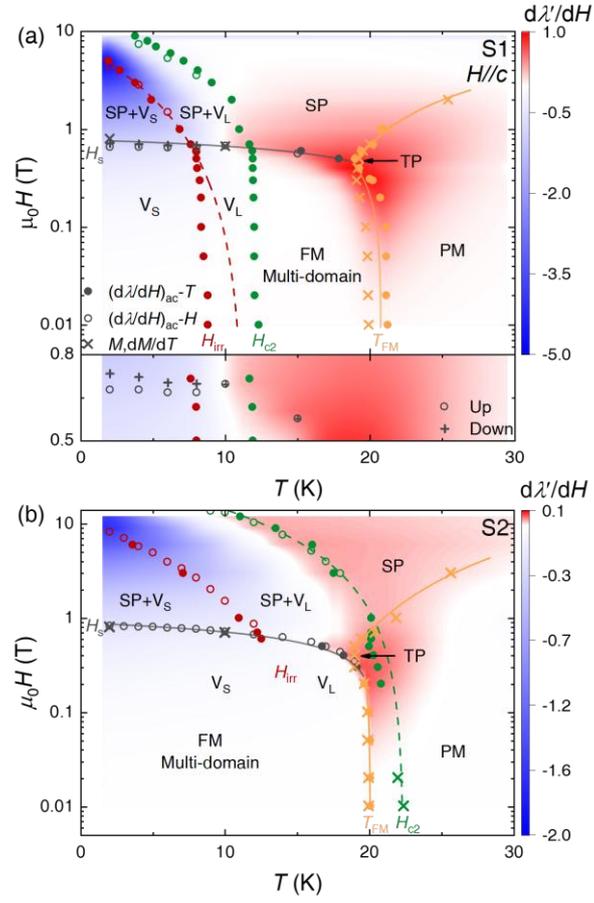

FIG. 5. Phase diagrams for (a) S1 and (b) S2, constructed from $(d\lambda/dH)_{ac}$ and magnetization data. Background is a contour map of $(d\lambda/dH)_{ac}$-$T$. Solid and dash lines are guides to the eyes.

# Supplemental Material to
# Distinct Suppression Mechanisms of Superconductivity by Magnetic Domains and Spin Fluctuations in EuFe$_2$(As$_{1-x}$P$_x$)$_2$ superconductors

Mengju Yuan[1,*], Nan Zhou[2,*], Ruixia Ti[3,*], Long Zhang[1], Chenyang Zhang[4], Tian He[5], Deliu Ou[1], Jingchun Gao[1], Mingquan He[1], Aifeng Wang[1], Junyi Ge[5,†], Yue Sun[2,‡], Yisheng Chai[1,§]

[1]*Low Temperature Physics Laboratory, College of Physics & Center of Quantum Materials and Devices, Chongqing University, Chongqing 401331, China*

[2]*School of Physics, Southeast University, Nanjing 211189, China*

[3]*School of Physics and Electronic Engineering, Xinxiang University, Xinxiang 453003, China*

[4]*School of Chemistry and Materials Engineering, Xinxiang University, Xinxiang 453003, China*

[5]*Materials Genome Institute, Shanghai University, Shanghai, 200444, China.*

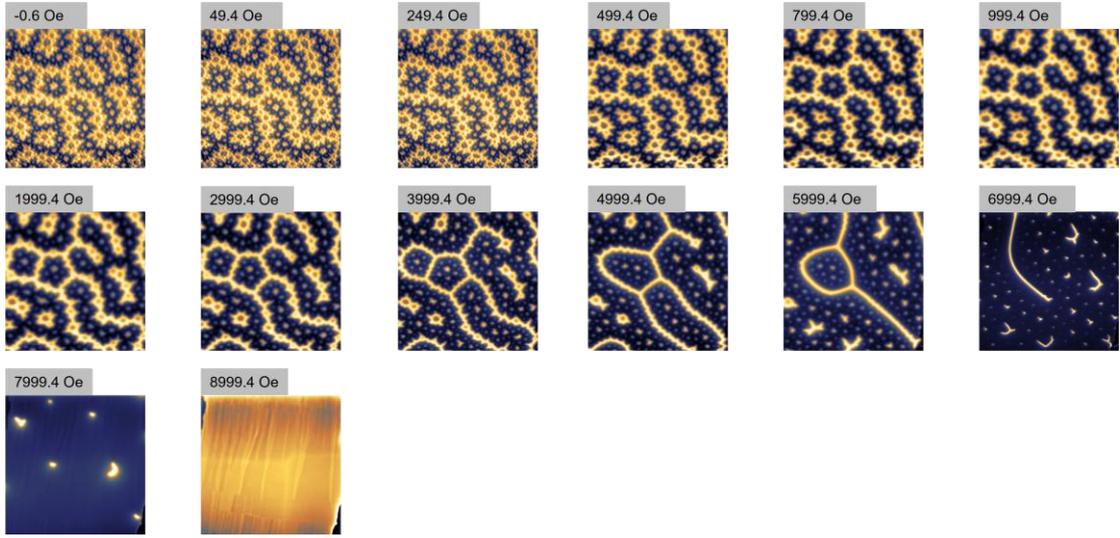

FIG. S1. Magnetic force microscopy images showing the evolution of magnetic patterns after FC at various magnetic fields and 2 K. The scanned area for each image is 18×18 μm$^2$.

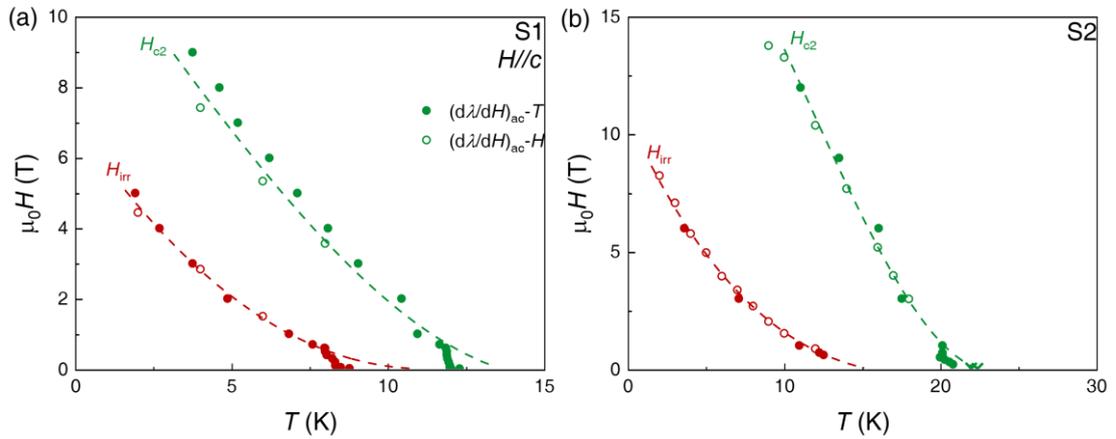

FIG. S2. Irreversibility field $H_{irr}$ and upper critical field $H_{c2}$ for sample (a) S1 and (b) S2 as a function of temperature in linear coordinates. The dashed lines are guided by extending the high-field evolution of $H_{irr}$/$H_{c2}$ into the low-field region.